\documentclass[11pt]{article}
\usepackage{moriond,epsfig}

\newcommand\kt{{k_\perp}}

\begin{document}
\begin{flushright}
  RAL-TR-1998-047 \\ hep-ph/9807226
\end{flushright}
\vspace*{3.4cm}
\title{JETS IN HADRON COLLISIONS IN QCD}

\author{MICHAEL H. SEYMOUR}

\address{Rutherford Appleton Laboratory, Chilton, Didcot, Oxfordshire.  
OX11 0QX.  England.}

\maketitle\abstracts{
\pretolerance 800
I briefly discuss three topics related to the hadroproduction of jets:
jet definitions; jet structure; and the underlying event.}

\section{Introduction}

Jets of hadrons are one of the most striking features of hadronic
collisions.  They also provide much of the interesting physics, both
directly in QCD studies, and indirectly in reconstruction of particles
that decay to jets, for example the top quark.  In recent years, we have
grown increasingly optimistic that jet physics is entering a new era of
`precision measurements', based on the increasing precision of
perturbative calculations and parton distribution functions on the one
hand, and the experimental data on the other.  In this talk, I would
like to discuss three potential flies in the ointment~-- areas in which
soft or semi-soft physics might spoil the connection between the hard
parton-level calculations and the hadron-level data.

It has recently been realized that the jet definition in current use,
the Snowmass-inspired iterative cone algorithm, is not infrared safe.
This means that we cannot calculate jet cross sections perturbatively.
In section~\ref{defs} I briefly discuss why not, what this might mean,
and how we can rectify it.

Modelling the internal structure of jets well is important, because this
determines the dependence of jet cross sections on the jet definition
and, to some extent, the systematic errors on jet measurements.  Direct
measurement of the jet structure is therefore important as a cross-check
of this modelling, as well as as a test of QCD in its own right.  In
section~\ref{shapes} I discuss the measurements that have been made to
date, and prospects for improved measurements using better observables.

One of the biggest uncertainties in jet physics is the correction due to
the `underlying event', the soft collision undergone by the hadron
remnants in addition to the hard parton scattering that produces the
jets.  In section~\ref{ue} I discuss why this is so important, and how
we can constrain current models better.

Finally in section~\ref{conc} I draw some conclusions.

\section{Jet definitions}
\label{defs}

The `Snowmass Accord'[\ref{snowmass}] was the first serious attempt to
standardize jet definitions across different experiments and between
theory and experiment.  It defined a jet as a direction that maximizes
the amount of transverse energy flowing through a cone centred on its
direction.  The transverse energy, $E_T$ of the jet is simply the scalar
sum of the transverse energies of the particles in it and its direction
in pseudorapidity, $\eta$, and azimuth, $\phi$, space is given by the
$E_T$-weighted averages of the particles in it.

However, this definition suffers from two main problems, the solution of
which have lead to more-or-less each experiment using a slightly
different definition.  Firstly, a global maximization is far too costly
in computer time, so a local maximization is used, starting from a
`seed' direction.  Typically, every calorimeter cell above a given $E_T$
is used as a seed.  Secondly, after doing this maximization, several
jets can be overlapping, sharing some particles in common.  The way this
is resolved depends on the amount of shared energy: if it is above some
threshold, the jets are merged together; and if it is below, they are
split, with the particles being shared out according to which jet they
are nearest.  The precise details of the versions used by
CDF[\ref{CDFcone}] and D0[\ref{D0cone}] are summarized in
Ref.~[\ref{Sshape}].

\begin{figure}[thbp]
\begin{minipage}[t]{.47\linewidth}
  \centerline{
    \resizebox{!}{5.5cm}{\includegraphics{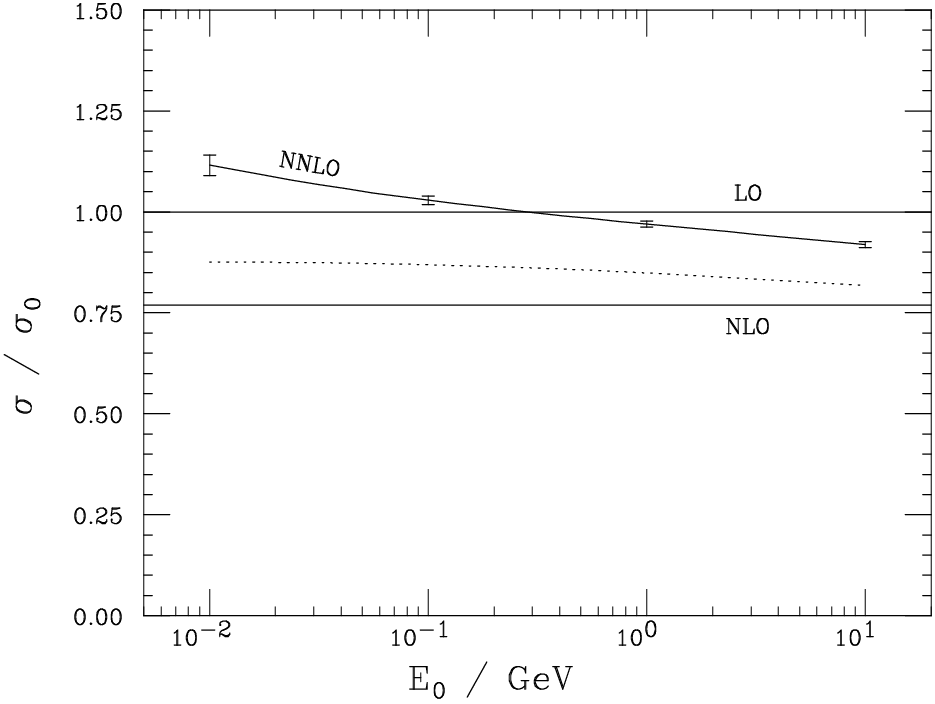}}
    }
  \caption[]{{\it The seed cell threshold dependence of the inclusive
      jet cross section in the D\O\ jet algorithm with $R=0.7$ in
      fixed-order (solid) and all-orders (dotted) calculations.  Taken
      from~[\ref{Sshape}].}}
  \label{Edep}
\end{minipage}
\hspace{0.78cm}
\begin{minipage}[t]{.47\linewidth}
  \centerline{
    \resizebox{!}{5.5cm}{\includegraphics{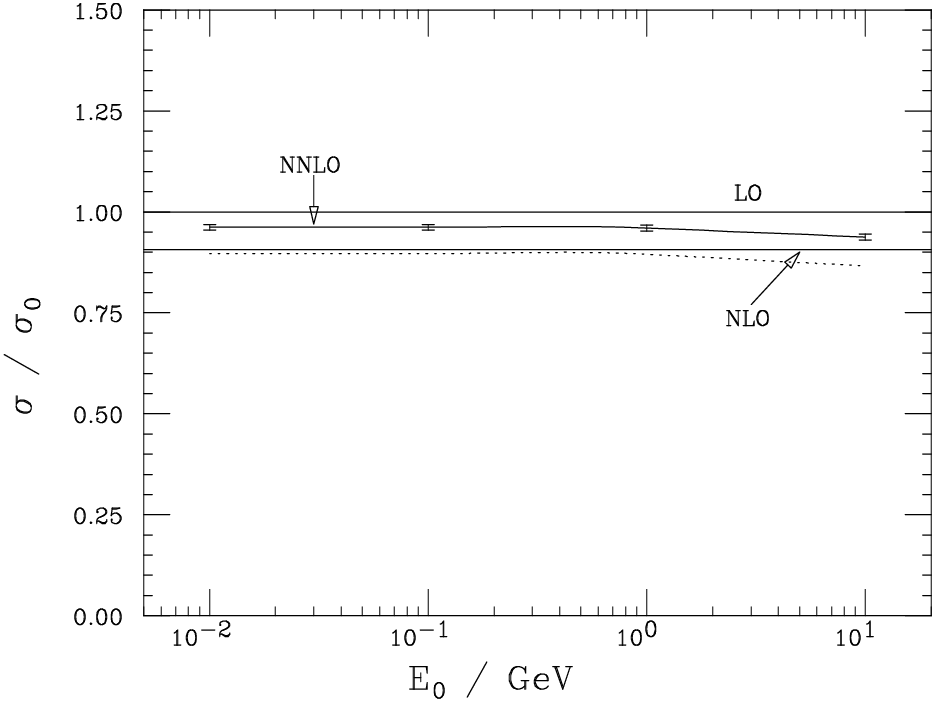}}
    }
  \caption[]{{\it As Fig.~\ref{Edep}, but in the improved iterative cone
      algorithm, in which midpoints of pairs of jets are used as
      additional seeds for the jet-finding.  Taken
      from~[\ref{Sshape}].}}
  \label{improved}
\end{minipage}
\end{figure}

Unfortunately, it turns out that the precise definition of the seeds
makes the merging/splitting step infrared-unsafe.  In order to calculate
cross sections in perturbation theory, the jet definition must be
insensitive to the presence of infinitely soft gluons.  This is not the
case in the algorithm as defined above, which is manifested as a
logarithmic dependence on the seed cell energy threshold.  If this is
set to zero, corresponding to an infinitely good detector, the cross
section is infinite at NNLO, as shown in Fig.~\ref{Edep}.
This problem had not previously been noticed because it is first
manifested in four-parton final states.  Exact order-by-order
calculations for two-jet cross sections have only reached NLO,
corresponding to three-parton final states.  However, a partial
calculation of the three-jet cross section to NLO, corresponding to
four-parton final states, has recently been completed[\ref{KGthree}],
and a negative-infinite cross section was found.  This is a direct
reflection of the positive-infinite inclusive two-jet cross section I
found, as the sum of the two is finite.  Furthermore, as explained in
Ref.~[\ref{Sshape}] Sudakov suppression prevents this effect from being
visible in the all-orders result (also shown in Fig.~\ref{Edep}), so
this problem was not identified using parton shower Monte Carlo programs
either.

A simple solution is to add a new step to the clustering process: after
iterating from the seed cells but before resolving overlapping jets,
consider the midpoint of every pair of jets found so far as a seed
direction and iterate again.  As shown in Fig.~\ref{improved}, this
solves the problem and gives a cross section that is finite in any order
of perturbation theory.  In my simplified approximation to higher
orders, it also gives a smaller correction from one order to the next,
but this will not necessarily be the case in the full calculation.

Despite this making the cross section finite, the fact that it makes no
difference at NLO means that the full dynamics are not well described by
NLO calculations.  A better solution is to use an algorithm in which
there is no merging/splitting ambiguity, such as the $\kt$
algorithm[\ref{CDSWkt},\ref{ESkt}].  As shown in Ref.~[\ref{ESkt}], this
is extremely similar to the iterative cone algorithm at NLO, provided
one sets the radius-like parameter about 1.35 times bigger in the $\kt$
algorithm than in the cone algorithm.  They also studied the
distribution of energy within the jet and found that more is
concentrated near the centre in the $\kt$ algorithm, meaning that it
`pays more attention' to the hard core of the jet and less to the soft
tails.  This is probably why it is superior to the cone for
reconstructing the masses of particles decaying to jets like the top
quark and Higgs boson[\ref{Ssearches}].

\section{Jet structure}
\label{shapes}

In leading order perturbation theory, jets are infinitely narrow.  A NLO
calculation of a jet cross section gives the first non-zero contribution
to the width.  At higher perturbative orders this width is increased
by multiple gluon emission, and ultimately by non-perturbative
hadronization effects.  Therefore by studying the internal structure of
jets, we are probing the details of the confinement mechanism by which a
hard parton evolves to a jet of hadrons.  This picture is somewhat
complicated by the underlying event, as I will discuss more in the next
section.

The simplest probe of jet structure is the jet shape, essentially a map
of how the jet's energy is distributed as a function of angle from its
centre.  The first perturbative calculation of the jet shape was made in
Ref.~[\ref{EKSshape}].  It was found that it does not describe the
data[\ref{CDFshape},\ref{D0shape}] very well at all, which was
attributed to the merging/splitting problems I mentioned earlier.  To
get around this, an adjustable parameter ``$R_{sep}$'' was introduced to
simulate the effect of higher orders in a NLO calculation.  More
recently, and particularly with the advent of good quality data from
HERA[\ref{ZEUSshape}], it was found that $R_{sep}$ had to vary as a
function of essentially all the kinematic variables available in order
to fit the data, thus losing all predictivity[\ref{KKshape}].  Since
these are only LO calculations of the jet structure, it is perhaps not
surprising that they do not do very well, particularly since it was
already known that their scale-dependence was large.

\begin{figure}[thbp]
\begin{minipage}[t]{.47\linewidth}
  \centerline{
    \resizebox{!}{5.5cm}{\includegraphics{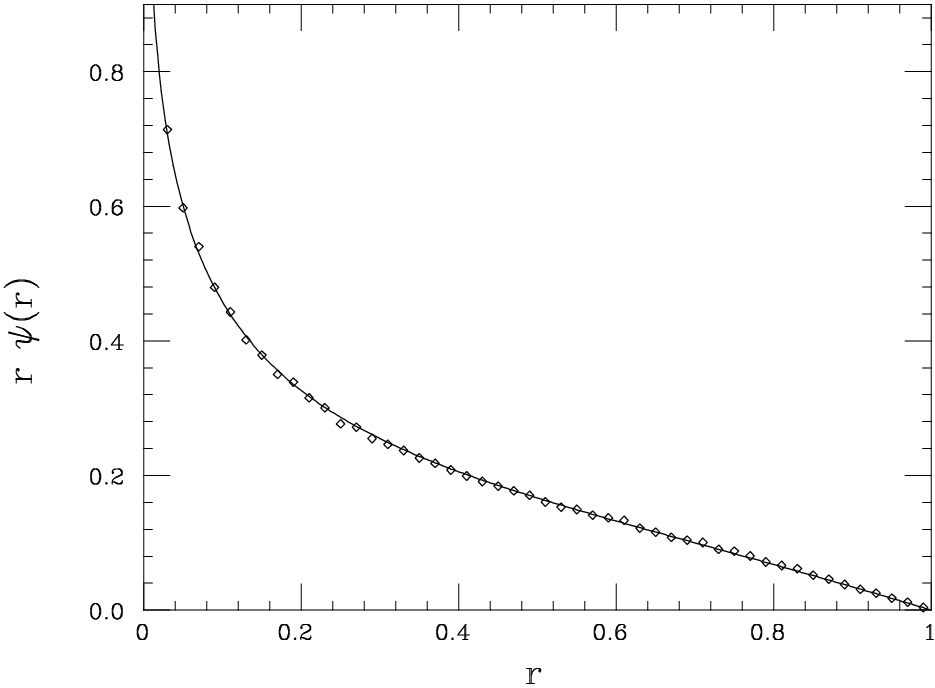}}
    }
  \caption[]{{\it The jet shape at leading order in the $\kt$ algorithm
      for a 50~GeV jet at $\eta=0$ according to the exact tree-level
      matrix elements (points) and the MLLA formula (curve).  Taken
      from~[\ref{Sshape}].}}
  \label{LO}
\end{minipage}
\hspace{0.78cm}
\begin{minipage}[t]{.47\linewidth}
  \centerline{
    \resizebox{!}{5.5cm}{\includegraphics{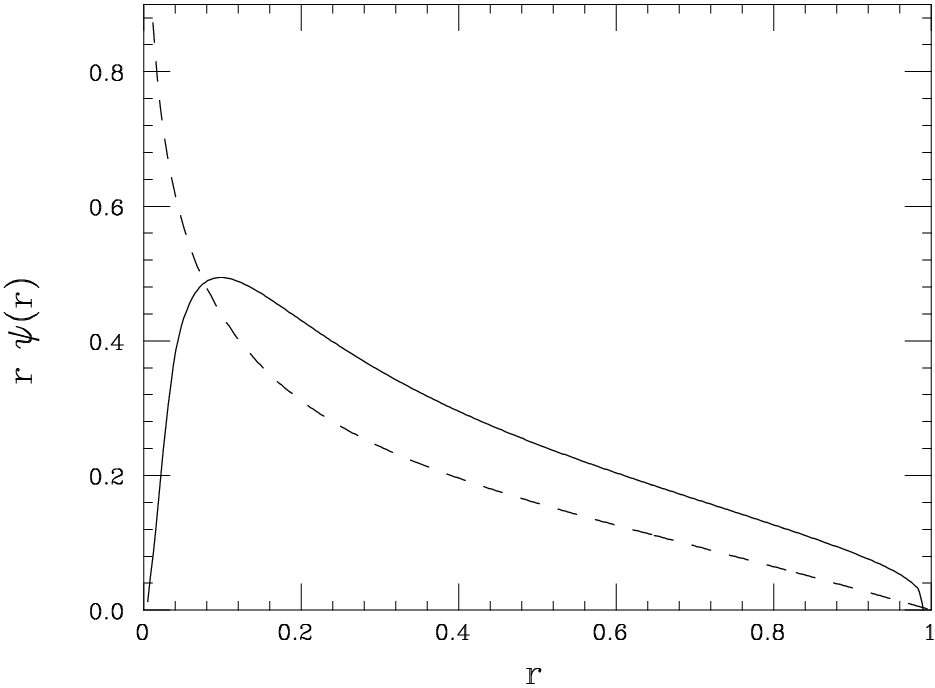}}
    }
  \caption[]{{\it Total effect of running coupling, power corrections
      and resummation on the shape of a 50~GeV jet in the $\kt$
      algorithm: LO (dashed) and with everything (solid).  Taken
      from~[\ref{Sshape}].}}
  \label{final50}
\end{minipage}
\end{figure}

In Ref.~[\ref{Sshape}], I used a simple model\footnote{A comparison of
  the model with the exact matrix elements is shown in Fig.\ref{LO},
  where it can be seen to be essentially perfect.} of the NLO matrix
elements to estimate the effect of higher order corrections, resummation
of large logs to all orders, and hadronization.  The conclusion, shown
in Fig.~\ref{final50}, is that these effects combined roughly double the
amount of energy near the edge of the jet, and give a dramatic overall
change in shape.
We should therefore not be too surprised that the LO calculations fail.
On the contrary, we should treat this as an indication that we can
really learn some physics from the jet shape.

A great deal of progress has recently been made in understanding power
corrections to semi-inclusive jet quantities in $\mathrm{e^+e^-}$
annihilation and DIS where the primary jets are principally quarks.  Jet
structure at hadron colliders offers a unique opportunity to make
similar studies of gluon-dominated jets and first studies in this
direction were made in Refs.~[\ref{Sshape},\ref{GGKshape}].

In comparison with Monte Carlo models, the results of
Refs.~[\ref{CDFshape},\ref{D0shape}] show that HERWIG's jets are always
too narrow, by an amount that varies with the jet kinematics.  Given
that the description of jets in $\mathrm{e^+e^-}$ annihilation is almost
perfect (at least on the scale of the disagreement in hadron
collisions), it is surprising that the description is so poor.  Asking
what is new in hadron collisions relative to $\mathrm{e^+e^-}$, three
features come to mind: gluon jets rather than quark jets; initial state
radiation; and the underlying event.  As shown by OPAL[\ref{OPALcone}],
one can implement cone algorithms in $\mathrm{e^+e^-}$ annihilation and
use vertex tagging to isolate a very pure sample of gluon jets.  These
are well-described by the parton shower models.  CDF made a very
detailed study of the distribution of the softest jet in three-jet
events[\ref{CDFcoherence}], which is very sensitive to the treatment of
initial-state radiation, and even interference between initial- and
final-state.  This was also well-described by HERWIG.  Therefore we are
left with the underlying event as the most viable culprit, which I will
discuss more in the next section.

\begin{figure}[thbp]
  \begin{minipage}[t]{0.55\linewidth}
    \centerline{
      \resizebox{!}{5.5cm}{\includegraphics{moriond98_05.ps}}
      }
    \caption[]{{\it Leading order (dashed) and resummed results for the
        subjet multiplicity in a 100~GeV jet at the Tevatron.  Taken
        from~[\ref{Ssubjet}].}}
    \label{subjet1}
  \end{minipage}\hfill\begin{minipage}[t]{0.4\linewidth}
    \centerline{
      \resizebox{!}{5.5cm}{\includegraphics{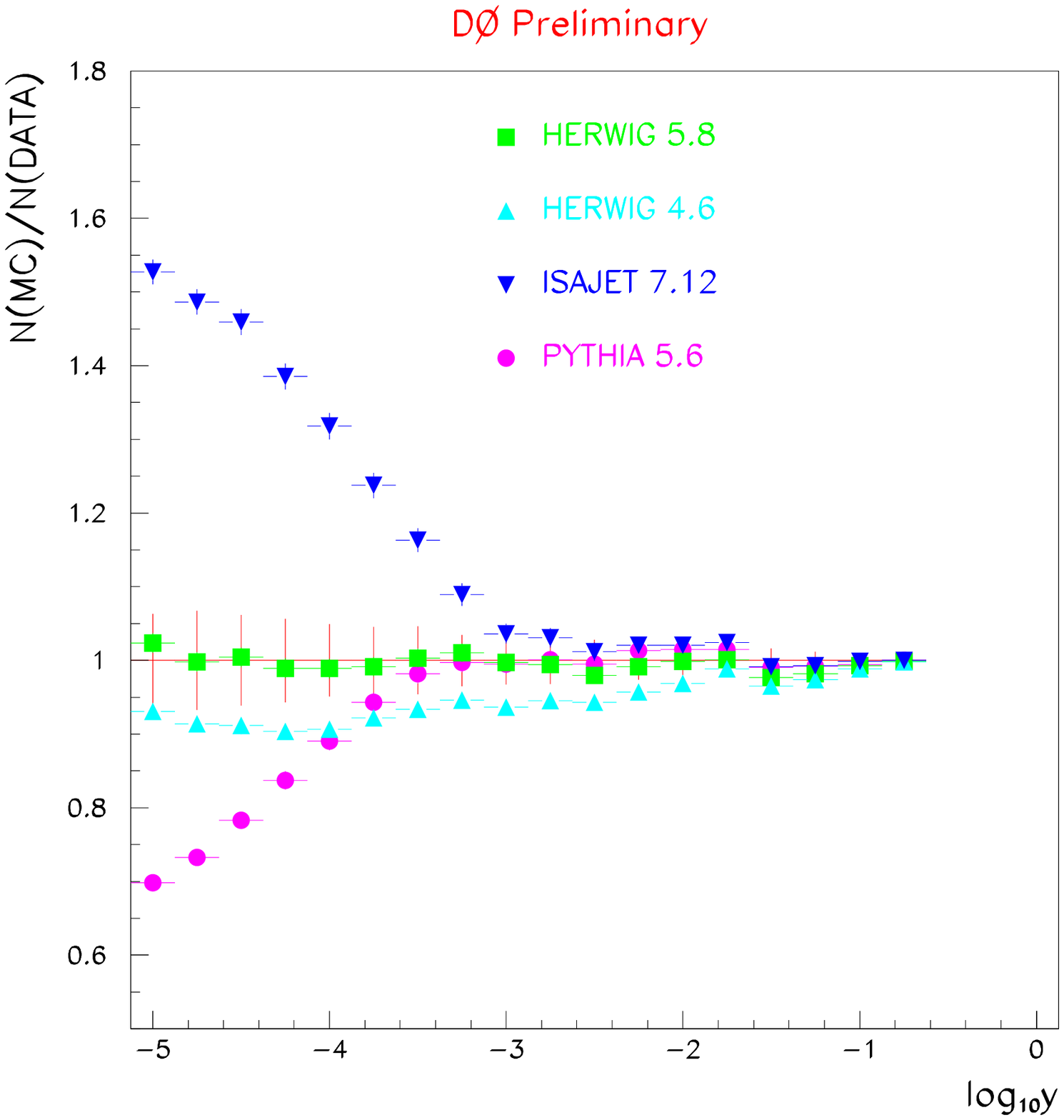}}
      }
    \caption[]{{\it The subjet multiplicity in a 300~GeV jet at detector
        level in various models, normalized to the data.  Taken
        from~[\ref{D0subjet}].}}
    \label{subjet2}
  \end{minipage}
\end{figure}

It is also known from $\mathrm{e^+e^-}$ annihilation that the energy
spread of a jet is not the best probe of its structure.  By resolving
the jet into subjets using a cluster algorithm such as the $\kt$, a much
more direct reflection of the underlying parton structure is obtained.
The simplest such quantity is the average number of subjets, calculated
in Ref.~[\ref{Ssubjet}].  The factorizing property of the $\kt$
algorithm means that leading and next-to-leading logs can be summed to
all orders, which is extremely important for small resolution criteria,
$y_{cut}$, as shown in Fig.~\ref{subjet1}.
So far, the subjet multiplicity has only been measured at detector
level[\ref{D0subjet}], only allowing comparison with Monte Carlo models.
There the agreement seen in Fig.~\ref{subjet2} is quite remarkable, at
least in HERWIG, given that at the left of the plot, 300~GeV jets are
being probed at a scale of only 1~GeV.

\section{The Underlying Event}
\label{ue}

To leading twist, final state properties in hadron collisions are
determined by the collision of a single parton from each hadron.
However, to give a complete description of the final state, one must go
beyond leading twist and consider how the rest of each hadron interacts
(producing the `underlying event').  This is an area of QCD that is
poorly understood both theoretically and experimentally, and a great
deal more study is warranted from both sides.  It is the source of the
majority of hadrons in an event and, although they are typically soft,
they have a major impact on jet physics.

At present we have two very different models for the underlying event:
`soft' and `hard'.  The soft model, implemented in HERWIG[\ref{herwig}],
treats the collision of the two remnants exactly like a minimum bias
event at the same energy, which is parametrized from UA5
data[\ref{UA5}].  The hard model, implemented in PYTHIA[\ref{pythia}]
and as an add-on package for HERWIG called JIMMY[\ref{jimmy}], uses
multiple perturbative scattering (`minijets') to produce the underlying
event.  Both give good descriptions of the average properties of events
and even correlations like the pedestal effect[\ref{SvZ},\ref{MWue}].
However, they give very different probabilities for rare upward
fluctuations in the amount of activity, which is what is most important
for determining jet cross sections.

The typical $E_T$ density produced by the underlying event is of order
1~GeV per unit $\eta\times\phi$, so one would na\"\i vely expect it to
have only a small effect on high $E_T$ jets.  This is not however the
case, because the perturbative jet spectrum falls so rapidly with $E_T$.
Therefore, although it is very rare for a jet to have its energy
increased by several GeV, at any given $E_T$ there are so many more jets
a few GeV below that the small fraction of them that get shifted up in
$E_T$ become a large fraction of the jets at that $E_T$.

It is predominantly the high $k_t$ tail of the underlying event
distribution that produces this shift, which is the feature that differs
most between the hard and soft models.  It would help to pin this high
$k_t$ tail down if the test proposed in Ref.~[\ref{MWue}] was carried
out.  Instead of just measuring the amount of $E_T$ in the tails of the
jet, the idea is to measure correlations between the $E_T$ on either
side of the jet, giving a probe of the `jetiness' of the pedestal.  The
proposal of Ref.~[\ref{MWue}] has been updated and adapted to
Tevatron-type analyses in Ref.~[\ref{Pue}], and it is hoped that it will
soon be measured.

The underlying event also has an important practical consequence for
Monte Carlo event generators because many events above a given $E_T$
come from partons at much lower $p_t$.  This means that the minimum
$p_T$ of scatters must be set much lower than the minimum jet $E_T$ in
order for it to have no effect, but then the efficiency is very low, as
most jets do not make it over the $E_T$ cut.  Unfortunately no solution
to this problem is foreseen at present.

\section{Conclusions}
\label{conc}

Despite the great advances made in jet physics in recent years, we
should not be over-complacent about our accuracy.  Non-perturbative
effects can be large, and must be understood before a connection can be
made between the precise parton-level calculations and the precise
hadron-level measurements.  The precision of both sides can be viewed as
an opportunity to learn more about QCD, but if on the other hand we
really want to test our perturbative calculations and measure $\alpha_s$
and the parton distribution functions, we need to constrain and improve
our models of the non-perturbative corrections.  Doing so will require
greater theoretical and experimental understanding.

\section*{References}
\begin{enumerate}
\item\label{snowmass}
  J.E. Huth {\it et al.}, in {\it Research Directions for the Decade},
  Proceedings of the Summer Study on High Energy Physics, Snowmass,
  Colorado, 1990, p.~134.
\item\label{CDFcone}
The CDF Collaboration, Phys.\ Rev.\ D45 (1992) 1448.
\item\label{D0cone}
The D0 Collaboration, {\em Fixed cone jet definitions in D0 and
  $R_{sep}$}, Fermilab preprint Fermilab-Pub-97-242-E
\item\label{Sshape}
  M.H. Seymour, Nucl. Phys. B513 (1998) 269
\item\label{KGthree}
  W.B. Kilgore and W.T. Giele, Phys.\ Rev.\ D55 (1997) 7183
\item\label{CDSWkt}
  S. Catani, Yu.L. Dokshitser, M.H. Seymour and B.R. Webber, Nucl.\
  Phys.\ B406 (1993) 187
\item\label{ESkt}
  S.D. Ellis and D.E. Soper, Phys.\ Rev.\ D48 (1993) 3160
\item\label{Ssearches}
  M.H. Seymour,  Z.\ Phys.\ C62 (1994) 127
\item\label{EKSshape}
  S.D. Ellis, Z. Kunszt and D.E. Soper, Phys.\ Rev.\ Lett.\ 69 (1992) 3615
\item\label{CDFshape}
  The CDF Collaboration, Phys.\ Rev.\ Lett.\ 70 (1993) 713
\item\label{D0shape}
  The D0 Collaboration, Phys.\ Lett.\ B357 (1995) 500
\item\label{ZEUSshape}
  The ZEUS Collaboration, Eur.\ Phys.\ J.\ C2 (1998) 61
\item\label{KKshape}
  M. Klasen and G. Kramer, Phys.\ Rev.\ D56 (1997) 2702
\item\label{GGKshape}
  W.T. Giele, E.W.N. Glover and D.A. Kosower, Phys.\ Rev.\ D57 (1998) 1878
\item\label{OPALcone}
  The OPAL Collaboration, Z.\ Phys.\ C63 (1994) 197
\item\label{CDFcoherence}
  The CDF Collaboration, Phys.\ Rev.\ D50 (1994) 5562
\item\label{Ssubjet}
  M.H. Seymour, Nucl.\ Phys.\ B421 (1994) 545
\item\label{D0subjet}
  R. Astur, in Proceedings of the 10th $\bar{p}p$ Workshop, Batavia,
  Illinois, 1995, p.~598
\item\label{herwig}
  G. Marchesini, B.R. Webber, G. Abbiendi, I.G. Knowles, M.H. Seymour
  and L. Stanco, Comput.\ Phys.\ Commun.\ 67 (1992) 465
\item\label{UA5}
  The UA5 Collaboration, Phys.\ Rept.\ 154 (1987) 247
\item\label{pythia}
  T. Sjostrand, Comput.\ Phys.\ Commun.\ 82 (1994) 74
\item\label{jimmy}
  J.M. Butterworth, J.R. Forshaw and M.H. Seymour, Z.\ Phys.\ C72 (1996) 637
\item\label{SvZ}
  T. Sjostrand and M. van Zijl, Phys.\ Rev.\ D36 (1987) 2019
\item\label{MWue}
  G. Marchesini and B.R. Webber, Phys.\ Rev.\ D38 (1988) 3419
\item\label{Pue}
  J. Pumplin, Phys.\ Rev.\ D57 (1998) 5787
\end{enumerate}

\end{document}